\documentclass[pra, twocolumn]{revtex4}

\usepackage{latexsym}
\usepackage{amssymb, amsmath, amsbsy, amsfonts, amssymb}
\usepackage{exscale}
\usepackage{bm, graphics}
\usepackage{graphicx, subfigure, pstricks}

\newcommand{\D}{\mbox{\rm d}}
\newcommand{\uhb}[1]{\underline{\hat{\mathbf{#1}}}}
\newcommand{\hb}[1]{\hat{\mathbf{#1}}}

\newcommand{\ket}[1]{$\left|#1\right\rangle$}

\usepackage{calrsfs}

\usepackage{tikz}
\usetikzlibrary{arrows}

\usepackage{xspace}
\newcommand*{\eg}{e.g.\@\xspace}
\newcommand*{\ie}{i.e.\@\xspace}

\begin{document}
\title{Three-dimensional cavity-assisted spontaneous emission as a
  single-photon source: Two cavity modes and Rabi resonance
}

\author{M. Khanbekyan}
\email[E-mail address: ]{khanbekyan@gmail.com}
\affiliation{Institut f{\"u}r Theoretische Physik, Universit{\"a}t
  Magdeburg, Postfach 4120, D-39016 Magdeburg, Germany}

\date{\today}

\begin{abstract}
Within the framework of exact quantum electrodynamics in dispersing
and absorbing media, we have studied the emission from an  initially
in the upper state prepared emitter in a high quality cavity in the
case, when there are two cavity modes that resonantly interact with
the emitter. 
In the case, when one of the modes is in resonance with the emitter
transition and the other mode is tuned to the frequency equal to the
Rabi splitting of the first mode, the effect of Rabi resonance is
observed. In particular, the second mode exhibits a substantial increase
of the one-photon Fock state efficiency and enhancement of the emission
spectrum in comparison to the exact resonant case.

\end{abstract}


\maketitle

\section{Introduction}
\label{introduction}

The interaction of a single emitter with the electromagnetic field
inside a high quality ($Q$) cavity has promising implications towards
realization of various schemes in quantum optics~\cite{reiserer:1379}
 and related fields
such as quantum information science~\cite{mabuchi:1372}. 
For scalable and integrable quantum information processing it is
furthermore vital to obtain high-efficiency quantum
entanglement. Various 
implementations in recent years include polarization entanglement of
photons by means of parametric down-conversion in nonlinear optical
crystals~\cite{kwiat:4337}, by using the biexciton (XX)--exciton (X)
cascade of single semiconductor quantum dots (QDs)~\cite{akopian:130501}, or via
time-bin entanglement~\cite{jayakumar:4251}.
Beyond that, the usage of
the strong coupling of the electromagnetic field with an emitter
placed in a high-finesse cavity promises the most substantial
improvement for both the efficiency of entanglement distribution
schemes and demonstration of violation of a Bell inequality at larger
scales  (see, \eg, Ref.~\cite{rauschenbeutel:050301}).  
Very widespread schemes of a strong coupling regime between a
single emitter and a single cavity mode in the domains of 
atomic~\cite{miller:551, wilk:488} and
semiconductor~\cite{khitrova:81} systems have been realized. 
Recently, increasing attention has been given to a new perspective of
quantum entanglement, namely, generation of multipartite quantum
$W$~states, that is, states exhibiting a uniform distribution of a
single photon across multiple electromagnetic field modes~\cite{tan:252,
  dunningham:180404}. 
Genuine $N$-partite entanglement states emerge as a resource for
exploring fundamental aspects of quantum theory 
(see, \eg, Refs.~\cite{aharonov:052108, dunningham:180404})
and
hold great promise for a number of
applications as quantum information protocols~\cite{nielsen} 
or robust quantum network
schemes (see, \eg, Refs.~\cite{fujii:050303, oezdemir:103003}).
In this context, various attempts have been made 
to realize an experiment that can unambiguously demonstrate the
nonlocality of a single particle~\cite{bjoern:180401, babichev:193601,
  dangelo:052114}.

The recent progress in semiconductor nanotechnology has substantially
facilitated the fabrication of various optical 
resonator structures characterized by a large confinement of light in
well-defined spatial and spectral mode profiles, enabling low
mode-volume, high quality factor emitter--cavity systems~\cite{reitzenstein:1733}. 
In particular, several experimental groups realized 
cavity QED (cQED) systems with large vacuum field amplitude at a QD 
placed in a high-$Q$ low volume microcavity~\cite{michler}.  
In the past decade the strong coupling regime of coupling of single
QDs with optical cavities has been realized in a number of high-quality
semiconductor micro- and nanostructures, as in photonic crystals~\cite{yoshie:200},
micropillars~\cite{reithmaier:197}, and microdisks~\cite{peter:067401}.
The interesting feature of micropillar cavities is the existence of two
spectrally separated orthogonal, linearly polarized high-$Q$
components of the fundamental optical mode. Here, the frequency splitting 
arises as the result of a slightly elliptical cross-section of 
the micropillar, which lifts the degeneracy of the resonator 
fundamental mode. 
It has been shown that these two modes can
couple with a single QD in the strong coupling
regime~\cite{reitzenstein:121306},
where temperature tuning has been used for variation of the transition
mode frequency.  
%
Moreover, the two polarization modes can couple
 with the common QD gain medium in micropillar
lasers~\cite{leymann:xxx}.
In particular, in the latter case the collective interaction of QDs
with the micropillar modes gives rise to an unconventional normal-mode
coupling between the modes~\cite{khanbekyan:043840}.
Then, spontaneous emission of a two-level atom resonantly interacting
with two modes of the cavity at the same time theoretically has been
discussed in Ref.~\cite{ho:85} within a simple generalized
Jaynes-Cummings approach. The model, however, does not allow the study
of the outgoing field spectra corresponding to the individual cavity
modes.



In earlier works, we have studied within the frame of exact quantum
electrodynamics~\cite{fidio:043822} the single-photon emission of a single 
emitter in a high-$Q$ cavity. In particular, we have shown the generation of 
single-photon wave packets with time-sym\-metric spatio-temporal  
shapes~\cite{khanbekyan:013822}, and
the generation of single-photons with the wave packets shorter than
the cavity decay time~\cite{fidio:043822}. More recently, we have
presented the feasibility of generation of one-photon Fock state wave
packets of desired shape with high efficiency by adjusting the shape
of the pump pulse applied to the three-level $\Lambda$-type
emitter~\cite{khanbekyan:013803}.
 
In the present article we extend the theory to the case of strong
coupling of a single two-level emitter with two orthogonal high-$Q$
polarization modes at the same time. 
The practical realization of simultaneous interaction of a single
emitter with two cavity modes can be realized, \eg, in semiconductor
systems~\cite{majumdar:183601, reitzenstein:235313}. In the case of
the micropillar cavities~\cite{reitzenstein:235313}, two orthogonal, linearly
polarized high-$Q$ components of the fundamental optical mode can be
spectrally separated due to slight asymmetry of the micropillar
cross-section. 
For a single QD interaction 
the QD exciton transition can be tuned by means 
of variation of the temperature. 
Here, we study, in particular, 
the quantum state of the excited outgoing field in detail.
To illustrate the theory we calculate quasiprobability distributions of
the individual polarization modes. The general expressions are derived
for the mode functions of the two polarization outgoing modes, which
allows us to study various coupling regimes of emitter--bimodal
cavity interaction.

The paper is organized as follows. The basic equations for the
resonant interaction of an emitter with a cavity-assisted
electromagnetic field are given in Sec.~\ref{sec2}. The theory is then
presented for the case of the resonant interaction with two polarization modes 
of high-$Q$ cavity. In Sec.~\ref{sec3}
 the Wigner function of the quantum state of the excited outgoing mode
 is studied. 
A summary and some concluding remarks are given in Sec.~\ref{sec9}.


\section{Basic equations}
\label{sec2}

\subsection{Quantization scheme}
\label{sec2.1}

Let us consider a single emitter (position ${\bf r}_A$)
that interacts with the electromagnetic field
in the presence of a dispersing and absorbing dielectric medium with a
spatially
varying and frequency-dependent complex permittivity 
\begin{equation}
    \label{1.0}
      \varepsilon({\bf r},\omega) = \varepsilon'({\bf r},\omega)
      + i\varepsilon''({\bf r},\omega),
\end{equation}
with the real and imaginary parts $\varepsilon'({\bf r},\omega)$
and $\varepsilon''({\bf r},\omega)$, respectively.
Applying the multipolar-coupling scheme in electric dipole
approximation, we may write the Hamiltonian
that governs the temporal evolution
of the overall system, which consists of the electromagnetic
field, the dielectric medium (including the dissipative degrees
of freedom), and the emitter coupled to the field,
in the form of~\cite{knoell:1, vogel}
\begin{equation}
   \label{1.1}
        \hat{H} =
        \hat{H}_\mathrm{field} + \hat{H} _\mathrm{A}
        + \hat{H} _\mathrm{int},
\end{equation}
where 
\begin{equation}
   \label{1.3}
        \hat{H}_\mathrm{field} =
        \int\! \D^3{r} \int_0^\infty\! \D\omega
      \,\hbar\omega\,\hb {f}^{\dagger}({\bf r },\omega)\cdot
      \hb{ f}({\bf r},\omega)
\end{equation}
is the Hamiltonian of the \mbox{field--me}\-dium system, where the
fundamental bosonic fields
\mbox{$\hb{ f}({\bf r},\omega)$}
and \mbox{$\hb{f}^\dagger({\bf r},\omega)$},
\begin{align}
    \label{1.5}
&      \bigl[\hb{f} ({\bf r}, \omega),
      \hb{f} ^{\dagger } ({\bf r }',  \omega ') \bigr]
      = 
      \delta (\omega - \omega  ')
      \bm{\delta}
      ({\bf r} - {\bf r }') ,
\\
\label{1.5-1}
&\bigl[\hb{f} ({\bf r}, \omega),
      \hb{f} ({\bf r }',  \omega ') \bigr]
= \bm{0},
\end{align}
play the role of the canonically conjugate system variables.
Further, 
\begin{equation}
 \label{1.7}
      \hat{H}_\mathrm{A} =
       \sum _m
        \hbar \omega _{m} \hat{S} _{mm}
\end{equation}
is the Hamiltonian of the emitter, where $\hat{S} _{mn}$ are the flip
operators, 
\begin{equation}
   \label{1.9}
   \hat{S} _{mn} =
   | m\rangle 
   \langle n |
,
\end{equation}
with \ket{m} being the energy eigenstates of the emitter. 
Finally, 
\begin{equation}
 \label{1.11}
         \hat{H} _\mathrm{int}=
        -
	\hb{ d}\cdot  \hb{E}({\bf r}_A)
\end{equation}
is the emitter-field coupling energy, where 
  \begin{equation}
    \label{1.12}
     \hb{ d} = \sum _{mn}
     {\bf d} _{mn}  \hat{S} _{mn}
 \end{equation}
is the electric dipole-moment operator of the emitter 
($ {\bf d} _{mn}$ $\!=$ $\!\langle m|
\hb{ d} | n \rangle$), and the operator of the
medium-assisted electric field $\hb{E}({\bf r})$
can be expressed in terms of the variables
$\hat{\mathbf{f}}(\mathbf{r},\omega)$ and
$\hat{\mathbf{f}}^\dagger(\mathbf{r},\omega)$ as
follows:
\begin{equation}
\label{1.13}
\hb{E}({\bf r}) = \hb{E}^{(+)}({\bf r})
        +\hb{E}^{(-)}({\bf r}),
\end{equation}
\begin{equation}
\label{1.15}
\hb{E}^{(+)}({\bf r}) = \int_0^\infty \D\omega\,
      \uhb{E}({\bf r},\omega),
\quad
\hb{E}^{(-)}({\bf r}) =
[\hb{E}^{(+)}({\bf r})]^\dagger,
\end{equation}
\begin{multline}
      \label{1.17}
      \uhb{ E}({\bf r},\omega) =
\\
i \sqrt{\frac {\hbar}{\varepsilon_0\pi}}\,
\frac{ \omega^2}{c^2}
      \int \D^3r'\sqrt{\varepsilon''({\bf r}',\omega)}\,
      \mathsf{G}({\bf r},{\bf r}',\omega)
      \cdot\hb{f}({\bf r}',\omega).
\end{multline}
In the above, the classical (retarded)
Green tensor $\mathsf{G}({\bf r},{\bf r}',\omega)$
is the solution to the equation
\begin{equation}
      \label{1.19}
      \bm{\nabla}
\times
    \bm{\nabla}\!  \times \mathsf{G}  ({\bf r }, {\bf r }', \omega)
      - \frac {\omega ^2 } {c^2} \,\varepsilon ( {\bf r } ,\omega)
      \mathsf{G}  ({\bf r}, {\bf r }', \omega)
      =  \bm{\delta} ^{(3)}  ({\bf r }-{\bf r }')
      \end{equation}
together with the boundary condition at infinity,
$\mathsf{G}({\bf r},{\bf r}',\omega)\to 0$ if
$|\mathbf{r}-\mathbf{r}'|\to\infty$, and defines the structure of the
electromagnetic field
formed by the presence of dielectric bodies.

\subsection{Two-level emitter in a two-mode cavity}
\label{sec2.3}

\begin{figure}[t!]
        \includegraphics[width=0.48\textwidth]{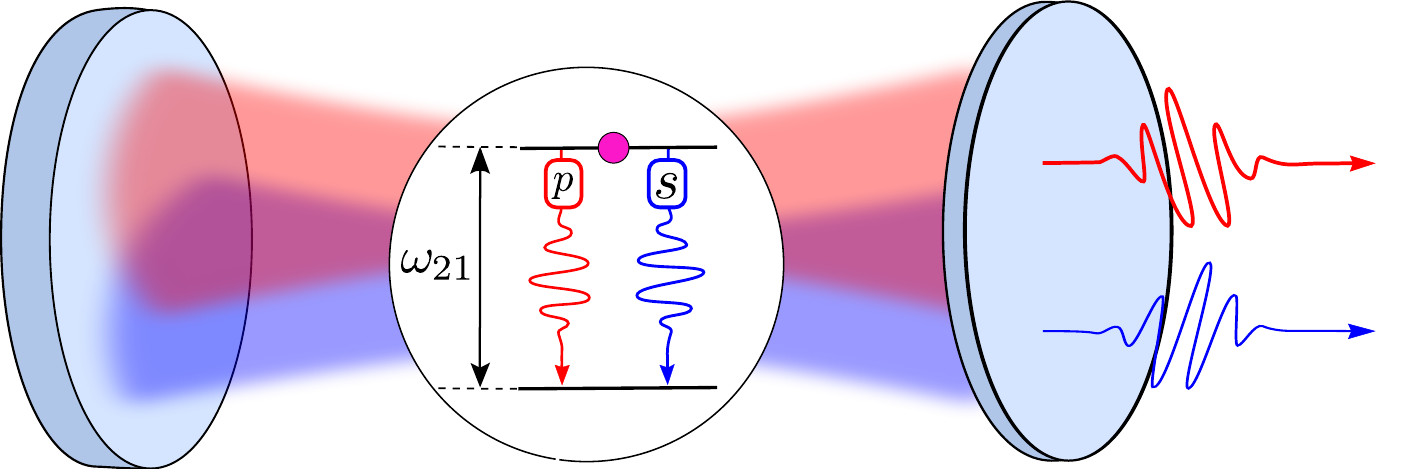}
	\caption{Scheme of a single emitter 
          (two-level system, initially prepared in the upper state)
          interacting with two modes with $s$- and $p$-polarization
          directions  in a high-$Q$ cavity. 
          The left mirror is perfectly reflecting and
          the right one is partially transparent. 
}
	\label{fig0}
\end{figure}
Let us focus on a single  atom-like emitter placed in a resonator
cavity and assume that only a single transition 
(\mbox{\ket{1} $\!\leftrightarrow$ \!\ket{2}}, frequency
$\omega_{21}$) 
 is quasi resonantly
coupled to a narrow-band cavity-assisted electromagnetic field, Fig.~\ref{fig0}. In
this case, the interaction Hamiltonian~(\ref{1.11}) in the
rotating-wave approximation reads 
\begin{align}
   \label{2.1}
        \hat{H}_\mathrm{int} =
& -
    i \sqrt{\frac {\hbar}{\varepsilon_0\pi}}\,
\int_0^\infty \D\omega\,
    \frac{ \omega^2}{c^2}
 \int\! \D^3r 
 	\sqrt{\varepsilon''({\bf r},\omega)}
 	\nonumber\\[1ex]&\,
 	\times
        {\bf d}_{21}\cdot
             \mathsf{G}({\bf r}_A,{\bf r},\omega)\cdot
\hb{f}({\bf r},\omega)
\hat{S}_{21}
            +
\mbox{H.c.}
\end{align}
In what follows we assume that the emitter is initially (at time $t$
$\!=$ $\!0$) prepared in the upper state \ket{2} and the rest of
the system, \ie, the part of the
system that consists of the
electromagnetic field and the dielectric media
(\ie, the cavity) is prepared in the ground state \ket{\{0\}},
defined by 
\mbox{$\hb{f}({\bf r},\omega)$  $\!$\ket{\{0\}} $\!=0$}. 
Since in the case
under consideration we may approximately span the Hilbert space of
the whole system by the single-excitation states, 
we expand the state
vector of the overall system at later times $t$ ($t\ge 0$) as
\begin{multline}
\label{2.3}
	|\psi(t)\rangle =
   C_2(t)
e^{-i\omega_{21} t}
  |\!\left\lbrace0\right\rbrace\!\rangle|2\rangle +
\\[.5ex]
+\!\int\! 
 \D^3r
\int_0^\infty \!\D \omega\,
   e^{-i\omega t}
   {\bf C}_1({\bf r}, \omega, t)\cdot
   \hb{f}^{\dagger}({\bf r}, \omega)
   |\!\left\lbrace0\right\rbrace\!\rangle |1\rangle ,
\end{multline}
where $\hb{f}^{\dagger}({\bf r}, \omega)\left|\{0\}\right\rangle$ is
an excited single-quantum state of the combined
field--cavity system.

It is not difficult to prove
that the Schr\"odinger equation for
\ket{\psi(t)}
leads to the following system of
differential
equations for the probability amplitudes
${\bf C}_1({\bf r}, \omega, t)$ and $C_2(t)$:
\begin{multline}
 \label{2.5}
 \dot {C}_2 =
      -\frac{1}{\sqrt{\pi \hbar \varepsilon _0  }}
      \int_0^\infty\! \D\omega\, \frac{\omega ^2}{c^2}
     \int \D^3r
      \sqrt{\varepsilon''({\bf r},\omega)}\,
\\[.5ex]
 \times
       {\bf d}_{21}\cdot
             \mathsf{G}({\bf r}_A,{\bf r},\omega)\cdot
      {\bf C}_1({\bf r}, \omega, t)
      e^{-i (\omega - \omega_{21})t},
\end{multline}
\begin{multline}
  \label{2.7}
  \dot {\bf C}_1({\bf r}, \omega, t) =
      \frac{1}{\sqrt{\pi \hbar \varepsilon _0 }}
      \frac{\omega ^2}{c^2}
       \sqrt{\varepsilon''({\bf r},\omega)}\,
\\[.5ex]
\times
      {\bf d}_{21}^*\cdot
             \mathsf{G}^*({\bf r}_A,{\bf r},\omega)
      C_2(t)
      e^{i (\omega - \omega_{21})t}.
\end{multline}
Further, substituting the formal solution to Eq.~(\ref{2.7}) 
[with the initial condition 
\mbox{$C_2(0)$ $\!=$ $\!1$} and \mbox{$ {\bf C}_1({\bf r}, \omega, 0)$ $\!=$ $\!0$}] 
into  Eq.~(\ref{2.5}) and using the integral relation 
 \begin{equation}
 \label{2.8}
 \frac {\omega ^2} {c^2}\!
 \int \! \D ^3 r'
 \varepsilon''({\bf r}', \omega)
 \mathsf{G}({\bf r},{\bf r}',\omega)
 \cdot
 \mathsf{G}^*({\bf r}'',{\bf r}',\omega)
 =
 \mathrm{Im}\,
 \mathsf{G}({\bf r},{\bf r}'',\omega)
 \end{equation}
we can derive the integro-differential equation
\begin{align}
 \label{2.9}
 \dot {C_2} =
\int_0^t \! \D t'\,
    K(t-t')
    C_2(t'),
\end{align}
where the kernel function $K(t)$ reads
\begin{multline}
  \label{2.11}
  K(t)=
     -\frac{1}{\pi \hbar \varepsilon _0  }
   \int_0^\infty\! \D\omega\, \frac{\omega ^2}{c^2}
    e^{-i (\omega - \omega_{21})t}
\\[.5ex]
 \times
   {\bf d}_{21}\cdot
            \mathrm{Im}\,  \mathsf{G}({\bf r}_A,{\bf r}_A,\omega)\cdot
             {\bf d}_{21}^*
.
\end{multline}
Having solved Eq.~(\ref{2.9}) for $C_2(t)$ we may eventually calculate 
${\bf C}_1({\bf r}, \omega, t)$ according to Eq.~(\ref{2.7})
\begin{multline}
  \label{2.21}
  {\bf C}_1({\bf r}, \omega, t)
  =
      \frac{1}{\sqrt{\pi \hbar \varepsilon _0 }}
      \frac{\omega ^2}{c^2}
      {\bf d}_{21}^*\cdot
             \mathsf{G}^*({\bf r}_A,{\bf r},\omega)
\\[.5ex]
\times
       \int_0^t  \D t'\,
       \sqrt{\varepsilon''({\bf r},\omega)}\,
      C_2(t')
      e^{i (\omega - \omega_{21})t'}.
\end{multline}

Without loss of generality, we may assume that the cavity is formed as a stratified
system, \ie, whose material properties do not change throughout each
plane orthogonal to a fixed direction ($z$ axis) defined as the direction of the
cavity outgoing field, 
\mbox{$\varepsilon({\bf r},\omega)$ $\!\rightarrow$ $\varepsilon(z,\omega)$}. 
A stratified system admits separation into an angular spectrum
representation of $s$- and $p$-polarized fields which do never
mix~\cite{born}, and  
the (nonlocal part of the) Green
tensor
can be expressed as a two-dimensional Fourier integral
\begin{align}
  \label{3.1}
     \mathsf{G}({\bf r},{\bf r}',\omega)=
     \frac{1}{(2\pi)^2}
      \int\! \D^2 k\,
      e^{i {\bf k}\cdot({\bm \rho}-{\bm \rho}')}
      \mathsf{G}(z, z', {\bf k},\omega),    
   \end{align}
with \mbox{${\bf r}$ $\!=$ $\!(z, {\bm \rho})$}, and
\begin{align}
  \label{3.3}
&     \mathsf{G}(z, z', {\bf k},\omega) 
\nonumber\\[1ex]&
=\!
\sum_{\sigma = s,p}\!
     \left[
       {\bf e}^+_\sigma ({\bf k  }) 
       {\bm g}^+_\sigma (z, z', {\bf k},\omega)
       +
        {\bf e}^-_\sigma ({\bf k  }) 
        {\bm g}^-_\sigma (z, z', {\bf k},\omega)
     \right],
\end{align}
where 
$\sigma$ indicates two polarization directions ($s$ and $p$) 
with unit vectors ${\bf e}^+_\sigma$ and ${\bf e}^-_\sigma$ for
outgoing and incoming waves, correspondingly, with the orthogonality
relations
\begin{align}
  \label{3.13}
  {\bf e}^\pm _\sigma({\bf k  }) \cdot {\bf e}^{\pm *} _{\sigma '} ({\bf k  })
  = \delta _{\sigma    \sigma '} ,
\end{align}
\begin{align}
  \label{3.14}
  {\bm g}^\pm _\sigma({\bf k  }) \cdot {\bm g}^{\pm *} _{\sigma '} ({\bf k  })
  = 0, \quad \sigma \neq \sigma '.
\end{align}

The Green tensor $\mathsf{G}({\bf r},{\bf r}',\omega)$ determines
the spectral response of the resonator cavity. 
For a sufficiently high~$Q$, the excitation spectrum
effectively turns into quasi-discrete sets of lines of
midfrequencies $\omega_{k,l}$ and widths $\Gamma_{k,l}$ in $s$- and
$p$-polarization directions, correspondingly, according to the poles
of the Green tensor at the complex frequencies  
\begin{equation}
  \label{2.13}
    \Omega _{k,l} = \omega_{k,l}
      - {\textstyle\frac {1} {2}}i\Gamma _{k,l}.
\end{equation}
In the following we assume that two modes of
the cavity-assisted field, namely, the $k$th mode of the $s$-polarization
direction and the $l$th mode of the $p$-polarization direction, are
quasi resonantly coupled with the emitter transition  
\mbox{\ket{1} $\!\leftrightarrow$ \!\ket{2}}, and  
\begin{equation}
\label{2.15}
\Gamma_k, \Gamma_l, 
{\textstyle\frac{1}{2}}|\omega_{k} - \omega_{l}|
 \ll {\textstyle\frac{1}{2}}|\omega_{m} - \omega_{k,l}|,\quad m\neq k,l.
\end{equation}
To calculate the kernel function $K(t)$, Eq.~(\ref{2.11}), we may
assume that the cavity-assisted field that resonantly 
interacts with the emitter can be approximated by Lorentzian
functions corresponding to 
the $k$th and $l$th modes
\begin{align}
  \label{2.17} 
&
     4 \omega \sqrt{\frac{\mu_0}{\pi \hbar}}
     {\bf d}_{21}\cdot
     \mathrm{Im}\,  \mathsf{G}({\bf r}_A,{\bf r}_A,\omega)\cdot
     {\bf d}_{21}^*
\nonumber\\&\quad
      = \sum_{j=k,l}\alpha_j(\omega_j)  \frac{1}{\pi}
      \frac{\frac {1}{2} \Gamma _{j}}{|\omega - \Omega_j|^2}.
\end{align}
Then,
we note
that, within the approximation scheme used, the frequency integration
can be extended to $\pm\infty$. Thus, using Eq.~(\ref{2.17}) we derive 
\begin{align}
  \label{2.19}
  K(t)=
    -
    \frac{1}{4}
    \sum_{j=k,l}
   \alpha_j (\omega_j) \Omega_j
    e^{-i(\Omega_j - \omega_{21})(t-t')}.
\end{align}
From Eq.~(\ref{2.9}) together with Eq.~(\ref{2.19})
we can conclude
that
\mbox{$R_j$ $\!\equiv\sqrt{\alpha_j(\omega_j)\omega_j}$}
can be
regarded as vacuum Rabi frequencies of emitter--cavity mode
interactions for the $k$th and $l$th modes, correspondingly. 
%

\section{Outgoing two-mode field}
\label{sec3}

\subsection{Field operators  }
\label{sec3.1}

For the sake of transparency, let us restrict our attention to the
case when the cavity is embedded in free space.
Then, inserting 
Eq.~(\ref{3.1}) together with Eq.~(\ref{3.3})
into Eq.~(\ref{1.17}) we find
\begin{align}
  \label{3.9}
&
 \uhb{ E}({\bf r},\omega) =
\frac{1}{(2\pi)^2}
  \sum_{\sigma = s,p}
      \int\! \D^2 k\,
       e^{i {\bf k}\cdot{\bm \rho}}
\nonumber\\[1ex]&
\times
       \left[
          {\bf e}^+_\sigma ({\bf k  })
          \hat{E}^+_\sigma (z, {\bf k},\omega)
          +
          {\bf e}^-_\sigma ({\bf k  })
          \hat{E}^-_\sigma (z, {\bf k},\omega)
       \right]         ,
\end{align}
with 
\begin{align}
  \label{3.11}
&
     \hat{E}^\pm_\sigma (z, {\bf k},\omega)
     = 
     i \sqrt{\frac {\hbar}{\varepsilon_0\pi}}\,
     \frac{ \omega^2}{c^2}
\nonumber\\[1ex]&
\times
\int \!\D z' \!
      \sqrt{\varepsilon''(z',\omega)}\,
      {\bm g}^\pm_\sigma (z, z', {\bf k},\omega)
      \cdot
      \hb{f}(z', {\bm k},\omega).
\end{align}
where
\begin{equation}
  \label{3.12}
  \hb{f}(z, {\bm k},\omega) =  
  \int\! \D^2\! \rho  e^{-i {\bf k}\cdot{\bm \rho}}\hb{f}(z, {\bm \rho},\omega)
\end{equation}
[\mbox{$\hb{f}(z, {\bm \rho},\omega)$ $\!\equiv\hb{f}({\bf r},\omega)$}].
In the following, we are interested in the two
polarization directions of the outgoing field 
$\hat{E}^+_\sigma (0^+, {\bf k},\omega)$
at the point 
\mbox{$z$ $\!=$ $\!0^+$} 
on the axis of the cavity outgoing field
propagation right outside the cavity. 
Using Eq.~(\ref{3.11}) together
with Eq.~(\ref{1.5}) we find
\begin{align}
  \label{3.15}
&
   \bigl[\hat{E}^+_\sigma (0^+, {\bf k},\omega),
   \hat{E}^{+\dagger}_{\sigma '} (0^+, {\bf k}',\omega ')
       \bigr]
\nonumber\\&\quad
      = 
      c_\sigma^+ ({\bf k},\omega)
      \delta _{\sigma \sigma '}
      \delta (\omega - \omega  ')
      \bm{\delta}
      ({\bf k} - {\bf k }') ,
\end{align}
where
\begin{align}
  \label{3.17}
   c_\sigma^+ ({\bf k},\omega)
   =
   \frac{4 \pi \hbar}{\varepsilon_0}\frac{\omega ^2}{c^2}
   \int \D z' \varepsilon''(z',\omega)
   |{\bm g}^+_\sigma (0^+, z', {\bf k},\omega)|^2
  . 
\end{align}
From Eq.~(\ref{3.15}) we can see that bosonic outgoing field operators
can be introduced according to
\begin{align}
  \label{3.19}
   \hat{E}^+_\sigma (0^+, {\bf k},\omega)=
   \sqrt{c_\sigma^+ ({\bf k},\omega)}
   \hat{b}_\sigma ({\bf k},\omega),
\end{align}
with
\begin{align}
  \label{3.21}
   \bigl[\hat{b}_\sigma ({\bf k},\omega),
   \hat{b}_{\sigma '}^\dagger ({\bf k}',\omega')
   \bigr]
   = \delta _{\sigma \sigma '}
      \delta (\omega - \omega  ')
      \bm{\delta}
      ({\bf k} - {\bf k }').
\end{align}

Notice, at this point no assumption is made about the details of stratification of the resonator media, the geometry and dielectric properties of which
define the functions ${\bm g}^+_\sigma (0^+, z', {\bf k},\omega)$.

\subsection{Quantum state of the outgoing field}
\label{sec3.2}

To calculate the quantum state of the outgoing field we start from
the multimode characteristic functional 
\begin{align}
  \label{3.23}
&
   C_\mathrm{out}[\beta_\sigma({\bf k},\omega),t] 
= \left\langle \psi(t)
  \right |
\nonumber\\&\,
 \times \exp\!\left[
    \sum _\sigma
  \int _0^\infty \!\D \omega\!
  \int\! \D ^2 k
  \beta _\sigma({\bf k}, \omega )
  \hat{b}_\sigma^\dagger({\bf k}, \omega )
 - \mathrm{H.c.}\right]\!
 \left| \psi(t) \right \rangle .
\end{align}
Applying the Baker--Campbell--Hausdorff formula
and recalling the commutation relation (\ref{3.21}),
we may rewrite $C_{{\rm out}}[\beta_\sigma({\bf k},\omega),t]$ as
\begin{multline}
\label{3.24}
 C_\mathrm{out}[\beta_\sigma({\bf k},\omega),t] =
\exp\!\left[ -{\textstyle\frac{1}{2}}\sum _\sigma\!
\int_0^{\infty}\!  \D\omega\!
\int\! \D ^2 k
      |\beta _\sigma({\bf k}, \omega )|^2
    \right]
\\[.5ex]\hspace{-4ex}\times
          \langle\psi(t)|
         \exp\!\left[
           \sum _\sigma\!
         \int_0^{\infty}\!
        \D\omega\!
        \int\! \D ^2 k
        \beta _\sigma({\bf k}, \omega ) \hat{b}_\sigma^{\dagger}({\bf k},\omega)
      \right]
\\[.5ex]\times
 \exp\!\left[
   -\! \sum _\sigma\!
         \int_0^{\infty}\!
        \D\omega\!
 \int\! \D ^2 k
        \beta^*_\sigma({\bf k}, \omega )  \hat{b}_\sigma({\bf k},\omega) 
      \right]
        \!
       |\psi(t)\rangle
        .
\end{multline}
To evaluate $C_\mathrm{out}[\beta_\sigma({\bf k},\omega),t] $ for the state 
\ket{\psi (t)} given by Eq.~(\ref{2.3}) we first note  
that from the
commutation relation (\ref{1.5}) together with the relation
$\hb{ f}(z, {\bf k},\omega)|\lbrace 0\rbrace\rangle$
$\!=$ $\!0$ it follows that
\begin{align}
  \label{3.25}
  \hb{ f}(z, {\bf k},\omega) \left| \psi(t) \right \rangle
   = 
   {\bf C}_1(z, {\bf k}, \omega, t)
   e^{-i\omega t}
   |1\rangle\
   |\lbrace 0\rbrace\rangle,
\end{align}
where
\begin{equation}
  \label{3.26}
     {\bf C}_1(z, {\bf k}, \omega, t)
     =\frac{1}{(2\pi)^2}  \int\! \D^2\! \rho
     e^{-i {\bf k}\cdot{\bm \rho}}
     {\bf C}_1({\bf r}, \omega, t).
\end{equation}
Hence, on recalling Eqs.~(\ref{3.11}) and (\ref{3.19}),
it can be shown that
\begin{equation}
  \label{3.27}
  \hat{b}_\sigma ({\bf k},\omega)
   |\psi(t)\rangle =
    F^*_\sigma({\bf k},\omega, t)
   |1\rangle\
   |\lbrace 0\rbrace\rangle,
\end{equation}
where
\begin{multline}
  \label{3.29}
    F_\sigma({\bf k},\omega, t)=-i
   \frac{1}{\sqrt{c^+_\sigma({\bf k},\omega)}}
   \sqrt{\frac{\hbar}{\varepsilon_0 \pi }}
    \frac{\omega^2}{c^2} e^{i\omega t}
\\[.5ex]
\times\!
      \int \!\D z
       \sqrt{\varepsilon''(z, \omega)}
       {\bm g}^{+*}_\sigma (0^+, z, {\bf k},\omega) \cdot
      {\bf C}_1^*(z, {\bf k}, \omega, t)
     ,
\end{multline}
with ${\bf C}_1(z, {\bf k}, \omega, t)$ being determined by
Eq.~(\ref{3.26}) together with Eq.~(\ref{2.21}). 
Then, combining	 Eqs.~(\ref{3.24}) and (\ref{3.27}) it is not difficult
to obtain
\begin{multline}
  \label{3.30}
  C_\mathrm{out}[\beta_\sigma({\bf k},\omega),t]
  =
\exp\!\left[ -{\textstyle\frac{1}{2}}\sum _\sigma\!
\int_0^{\infty}\!  \D\omega\!
\int\! \D ^2 k
      |\beta _\sigma({\bf k}, \omega )|^2
    \right]
\\[.5ex]\hspace{-4ex}\times
    \left[
      1-
\left|
\beta_s(t)+\beta_p(t)
\right|^2
\right],
\end{multline}
where
\begin{equation}
  \label{3.32}
  \beta_\sigma (t)= 
  \int_0^{\infty}\!  \D\omega\!\int\! \D ^2 k
  \beta _\sigma({\bf k}, \omega )
  F_\sigma({\bf k},\omega, t).
\end{equation}
As we can see from Eq.~(\ref{3.30}), the characteristic function of
the outgoing field 
does not factorize into a product of characteristic functions in $s$-
and $p$-polarization directions. 
The polarization modes are quantum correlated
in a nontrivial fashion, 
which shows the entangled nature of the quantum state under
consideration.

\subsection{Mode functions and Rabi Resonance}
\label{sec5}


%

The characteristic functions for the $s$- 
($p$)-polarization directions can be found from
$C_\mathrm{out}[\beta_\sigma({\bf k},\omega),t]$ by
simply setting, correspondingly, 
\mbox{$\beta_p({\bf k},\omega)$ $\!=$ $\!0$ } or
\mbox{$\beta_s({\bf k},\omega)$ $\!=$ $\!0$ }
in Eq.~(\ref{3.30}). 
Then, following Ref.~\cite{khanbekyan:013822},
we Fourier transform the characteristic functions for every
polarization direction and by choosing appropriate sets of
nonmonochromatic modes, we obtain  
the multimode Wigner functions of the quantum state
of the outgoing field in $\sigma$-polarization direction to be
\begin{equation}
\label{3.31}
W_{\sigma \mathrm{out}} (\gamma_{\sigma i}, t)
= 
W_{\sigma 1}(\gamma_{\sigma 1},t)
\prod_{ i\neq 1}
     W_{\sigma i}^{(0)}(\gamma _{\sigma i}, t),
\end{equation}
where
\begin{equation}
\label{3.33}
W_{\sigma 1}(\gamma,t)
= [1-\eta_\sigma(t)]W_{\sigma 1}^{(0)}(\gamma)
     +\eta_\sigma(t)W_{\sigma 1}^{(1)}(\gamma),
\end{equation}
with $W_{\sigma i}^{(0)}(\gamma)$ and $W_{\sigma i}^{(1)}(\gamma)$
being the Wigner functions of the vacuum
state and the one-photon Fock state, respectively, for the
$i$th nonmonochromatic mode of $\sigma$-polarization.
As we can see from Eq.~(\ref{3.31}), the nonmonochromatic modes with
\mbox{ $i$ $\!\neq$ $\!1$} are in the vacuum state. The modes labeled 
\mbox{ $i$ $\!=$ $\!1$}
with the nonmonochromatic mode functions
\begin{equation}
  \label{3.34}
    F_{\sigma 1}({\bf k}, \omega, t)
    = \frac{F({\bf k}, \omega , t)}{\sqrt{\eta_\sigma(t)}}
\end{equation}
are in the mixed state described by the Wigner functions 
$W_{\sigma 1}(\gamma,t)$.
Moreover, Eq.~(\ref{3.33}) reveals that $\eta_\sigma(t)$ defined by
\begin{equation}
 \label{3.35}
         \eta_{\sigma}(t)
=
\int_0^{\infty}\!\D\omega\,\!\int\! \D ^2 k\,
        |F_\sigma({\bf k},\omega ,t)| ^2
 \end{equation}
can be regarded as the efficiency to prepare the excited outgoing 
$\sigma$-polarization wave packet in a one-photon Fock state.

To calculate the nonmonochromatic mode functions 
$F_{\sigma 1}({\bf k},\omega ,t)$ we combine Eqs.~(\ref{2.21}) and
(\ref{3.29}) to derive 
\begin{multline}
  \label{5.1}
    F_{\sigma1}({\bf k},\omega , t)=
    -i\frac{1}{\sqrt{c^+_\sigma ({\bf k},\omega)\eta_{\sigma}(t)}}
    \frac{1}{\pi \varepsilon_0}
    \frac{\omega^2}{c^2}
             e^{i{\bf k}\cdot{\bm \rho}_A}
\\[.5ex]
\times\!
 	\int\!\D z
 	\varepsilon''(z, \omega)
 		{\bf d}_{21}\cdot
             \mathsf{G}(z_A,z,{\bf k},\omega)\cdot
             {\bm g}^{+*}_\sigma (0^+, z, {\bf k},\omega) 
       \\[.5ex]
\times\!      
             \int ^t _0 \D t'\, 
      C_2^*(t')e^{i\omega(t-t')}
     e^{i\omega_{21}t'}
   .
\end{multline}
As we have mentioned above, the Green tensor 
$\mathsf{G}({\bf r},{\bf r}',\omega)$, or in particular,  
${\bm g}^{+}_\sigma (0^+, z_A, {\bf k},\omega)$ 
and $\mathsf{G}(z_A,z,{\bf k},\omega)$
in Eq.~(\ref{5.1}),
determine the spectral resonances of the high-$Q$ cavity. 
In particular, the Green tensor terms define the $k$th and $l$th resonance frequencies
of the $s$- and $p$-polarization directions of the cavity, respectively
[complex frequencies $\Omega _{k,l}$, Eq.~(\ref{2.13})].  
Moreover, the first scalar product in Eq.~(\ref{5.1}),
\mbox{${\bf d}_{21}$ $\!\cdot$ $\!\mathsf{G}(z_A,z,{\bf k},\omega)$},
can be identified as the overlap of the emitter electric dipole 
moment and the cavity field. The second scalar product 
\mbox{$\!\mathsf{G}(z_A,z,{\bf k},\omega)$ $\!\cdot$ $\!{\bm g}^{+*}_\sigma (0^+, z, {\bf k},\omega)$}
describes the wanted radiative losses of the cavity modes  
$\gamma_{k,l}$ due to the transmission of the radiation through the
fractionally transparent mirror. 
%
%
It can be shown~\cite{khanbekyan:013803} that the efficiencies
$\eta_{\sigma}(t)$ are proportional to the ratio
$\gamma_{k,l}/\Gamma_{k,l}$, where the damping parameters
$\Gamma_{k,l}$ [see, \eg, Eq.~(\ref{2.13})] can be regarded as being the sum
$\Gamma_{k,l} = \gamma_{k,l} + \gamma_{k,l}^\prime$, where $\gamma_{k,l}^\prime$
describe the unwanted losses due to scattering and absorption, defined
by the complex permittivity $\varepsilon(z,\omega)$, Eq.~(\ref{1.0}).
%
Note, unwanted losses as low as $30\%$ have been reported in modern
high-$Q$ semiconductor cavities~\cite{ding:020401, somaschi:340},
which corresponds to 
\mbox{$\gamma_{k}/\Gamma_k$ $\!=\gamma_{l}/\Gamma_l$ $\!=0.7$}.

\begin{figure}[t!]
        \includegraphics[width=0.48\textwidth]{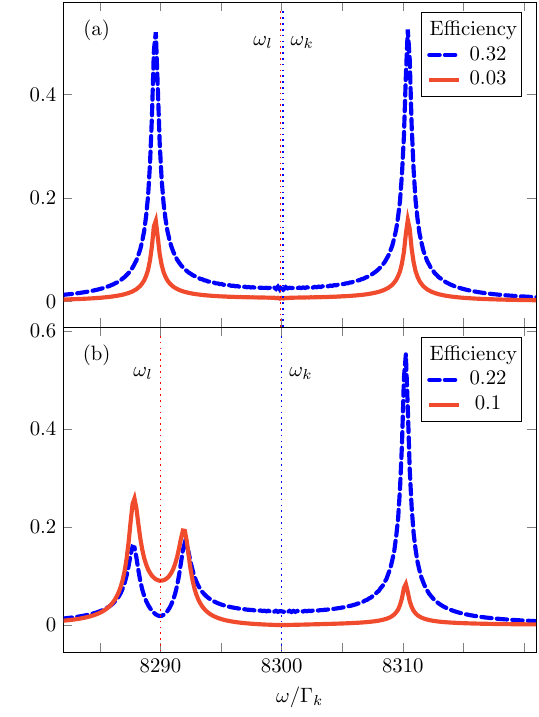}
	\caption{Spectral mode functions of the outgoing field 
          $|F_{\sigma1}({\bf k},\omega , T)|$ in 
		$s$-(blue, dashed) and $p$-(red, solid) polarization directions
                for ${\bf k}=0$, $\Gamma_l = 1.25\Gamma_k$,  $R_k = 20\Gamma_k$, 
                $R_l = 6\Gamma_k$,  
                $T = 10 \Gamma_k$,  
                $\gamma_{k} $ $\!=0.7\Gamma_k$,
                $\gamma_{l} $ $\!=0.875\Gamma_k$,
                $\omega_{21} =\omega_{k} =8300\Gamma_k$,
                and  
                (a) $\omega_{l} =8300\Gamma_k$, 
                (b) $\omega_{l} =8290\Gamma_k$.
}
	\label{fig1}
\end{figure}

Thus, to obtain the outgoing nonmonochromatic mode functions
$F_{\sigma 1}({\bf k},\omega , t)$ for the $s$- and $p$-polarization
directions, we first numerically solve the integro-differential Eq.~(\ref{2.9})
together with Eq.~(\ref{2.19}) and insert the solution for $C_2(t)$
into Eq.~(\ref{5.1}). To further evaluate Eq.~(\ref{5.1}),
we model the cavity as a planar dielectric multilayer system, where
the cavity body is confined by layers of perfectly reflecting and
fractionally transparent mirrors (see Ref.~\cite{khanbekyan:063812}
for details and Appendix~\ref{app} 
for the expression of the
corresponding Green tensor).   
%
%
%
The spectral mode functions 
%
$F_{\sigma 1}({\bf k},\omega , t)$ for the $s$- and $p$-polarization
directions
are plotted in Fig.~\ref{fig1} when the $k$th mode ($s$-polarization)
is in exact resonance with the emitter transition for the two
following cases.  
In the first 
case (Fig.~\ref{fig1}(a), mode $Q$ factors are $8300$ and $6640$ for
the $k$th and $l$th modes, correspondingly, which is in accordance with
the experimentally observed values~\cite{reitzenstein:235313}), the $l$th
mode ($p$-polarization) is also in resonance with the emitter transition.
Then, 
both mode spectra exhibit two peaks around the resonance frequency
featuring vacuum Rabi splitting of the emitter-field interaction.
Obviously, 
since the Rabi frequency $R_k = 20\Gamma_k$ is
higher than $R_l = 6\Gamma_k$,
the one-photon Fock state efficiency of the 
outgoing wave packet in $s$-polarization
is higher than the one of the $p$-polarization.
Further, in the second case, when the frequency of the $l$th mode 
coincides with the one of the peaks of the Rabi splitting of the
$k$th mode--emitter interaction [Fig.~\ref{fig1}(b), 
mode $Q$ factors are $8300$ and $6632$ for
the $k$th and $l$th modes, correspondingly], 
both modes spectra exhibit three-peak structures. 
Interestingly, by comparison of Figs.~\ref{fig1}(a) and \ref{fig1}(b) we notice,
that in the latter case the enhancement of the $p$-polarization
emission spectrum
and 
diminution of the 
$s$-polarization emission spectrum
is observed in comparison to the exact
resonance case.  
Moreover, one-photon Fock state efficiency of the 
$p$-polarization direction
is 
more than $3$~times larger in comparison to
the exact resonance case.

\section{Summary and concluding remarks}
\label{sec9}

Based on macroscopic QED in dispersing and absorbing media, we have
presented an exact description of the resonant interaction of a
two-level emitter in a high-$Q$ cavity with the cavity-assisted
electromagnetic field. In particular, we have studied the case when
two orthogonal closely-spaced polarization modes of the cavity at the
same time resonantly interact with the emitter. 
We have applied the theory to the determination of the Wigner function
of the quantum state of the excited outgoing field.  

Assuming that the Hilbert space of the total system is spanned by a
single-quantum excitation, 
we have shown that the polarization directions do not
factorize in the Wigner function of the outgoing
field. Thus, we may conclude that
the quantum state of the outgoing field represents
a nonseparable coherent superposition  of the two polarization
directions.   

To further exploit the nature of the resonant bimodal interaction, 
then we have studied quasiprobability distributions of the individual
polarization directions. 
In the case when both polarization modes are in exact resonance with the
emitter transition, the spectral mode functions of the two
polarization directions  
exhibit two-peak structures around the resonance
frequency. The single-photon Fock state efficiencies for 
every of the
polarization directions are determined by the corresponding Rabi
frequencies of the emitter-field coupling.
More interesting is the case when one of the modes is in resonance
with the emitter transition, and
the frequency of the second polarization mode is out of 
resonance but
coincides with the one of the peaks of
the Rabi splitting of the first mode.
In particular, in this case 
both spectra of the resonant and the off-resonant modes  
exhibit
three-peak structures. 
This is an example of the well-known Rabi resonance
scenario~\cite{bishop:105, grunwald:063806}, 
which shows 
that mode interaction in the bimodal cQED systems may drastically
change the features of the nonclassical radiation field emission. 
In particular,
as an important effect of the assisted Rabi resonance we observe at the
Rabi resonance frequency region drastic enhancement of the emission
spectrum of the off-resonant mode. At the same time
diminution of the spectrum of the resonant mode is observed. 
Moreover, the
one-photon Fock state efficiency of the off-resonant mode is substantially
larger in the case of the Rabi-resonance case in comparison to the
exact resonance case.    

In summary, spectral mode functions of the outgoing field of two
cavity modes simultaneously interacting with a single emitter in the
exact resonant case  exhibit similar 
two-peak structures featuring Rabi splitting of the strong
coupling.  
In the case, when the first mode is in resonance and the second mode is
tuned to the Rabi frequency of the coupling of the first mode, the
spectral mode functions of both cavity modes are split into triplets.

Here,
the polarization degree of freedom 
of the cavity modes
can be useful for mode-resolved spectroscopy of the outgoing 
field. 
This is important 
in view of a practical realization of the scheme,
both in the resonant case and in the Rabi resonance case. 
In the latter case, although the cavity modes are frequency separated
they both feature similar spectra with nearly identical peak structure. 
With respect to this case, another degree of freedom is necessary to
distinguish spectra of the individual modes. 
Therefore, to observe the effect of Rabi resonance in a practical
situation, another degree of freedom such as polarization is needed for the
mode-resolved spectroscopy of the outgoing field. Another possible
  realization of the simultaneous strong 
  coupling of two modes with a single emitter may be provided by 
  the interaction of transverse modes with a two-level atom, where
  mode-resolved spectroscopy is also possible, see, \eg,
  Ref.~\cite{wickenbrock:043817}. In Ref.~\cite{wickenbrock:043817},
  however, only the case when both modes are in exact resonance with
  the emitter is considered.





Finally, as an outlook, the system featuring tunable interaction of a
single emitter with two closely-spaced cavity modes can be used to
generate genuine multipartite polarization entangled states of
light. Quantification of the entanglement of the field modes in the
exact resonance and Rabi resonance scenarios is the subject of our
future investigations.

\begin{acknowledgments}
  The author acknowledges valuable discussions with
  \mbox{D.-G.~Welsch}, J.~Wiersig, and A.~Musia\l{}.
\end{acknowledgments}  

\appendix

\section{Green tensor for three-dimensional dielectric multilayer structure}
\label{app}
   The nonlocal part of the Green tensor for a three-dimensional dielectric multilayer structure reads 
   \begin{multline}
      \label{app:1}   
      \mathsf{G}
      	(z, z', \mathbf{k},  \omega )
\\
       =
      \frac {i} {2} \sum _{\sigma=s,p}\!
      \xi _\sigma \left[
      \bm{\mathcal{E}} ^{j>} _{\sigma}  (z, \mathbf{k},  \omega )\,
      \Xi^{jj'} _{\sigma}  \bm{\mathcal{E}} ^{j'<} _{\sigma}  (z', -\mathbf{k},  \omega)
      \Theta (z\!-\!z')
      \right.
 \\
      \left.
      +
      \bm{\mathcal{E}}^{j<} _{\sigma}  (z, \mathbf{k},  \omega )\,
      \Xi^{j'j} _{\sigma} \bm{\mathcal{E}} ^{j'>} _{\sigma} (z', -\mathbf{k},  \omega )
      \Theta (z'\!-\!z)
      \right] ,
      \end{multline}
where $z$ ($z'$) belongs to the layer $j$ ($j'$), and \mbox{$\xi _p$
  $\!=$ $\!1$}, \mbox{$\xi _s$ $\!=$ $\!-1$}. 
In the above, the functions
$\bm{{\mathcal{E}}} ^{j>} _{q} (\mathbf{k}, \omega, z)$
and $ \bm{{\mathcal{E}}} ^{j<} _{q} (\mathbf{k}, \omega, z)$
denote waves of unit strength, traveling, respectively,
rightward and leftward in the $j$th layer, and being reflected
at the boundary,
\begin{equation}
      \label{app:2}
\bm{\mathcal{E}} ^{(j)>} _{q}  (z, \mathbf{k},  \omega )
 = 
      \mathbf{e}^{(j)}_{q+} (\mathbf{k})  e^{i \beta _j (z-d _j)}
      + r^q _{j/n} \mathbf{e}^{(j)}_{q-} (\mathbf{k})
      e^{-i \beta _j (z-d _j)}, 
\end{equation}
\begin{equation}
      \label{app:3}
      \bm{{\mathcal{E}}} ^{(j)<} _{q}  
      (z, \mathbf{k},  \omega ) 
      =
      \mathbf{e}^{(j)}_{q-} (\mathbf{k})  e^{-i \beta _j z}
      + r^q _{j/0} \mathbf{e}^{(j)}_{q+} (\mathbf{k})  e^{i \beta _j
        z}, 
\end{equation}
and
\begin{equation}
      \label{app:4}
      \Xi^{jj'} _{q}
      =
      \frac{1}{\beta_n t^q_{0/n}}\,
      \frac{t^q_{0/j}e^{ i \beta _j d _j}}{D _{q j}}\,
      \frac{t^q_{n/j'}e^{ i \beta _{j'} d _{j'}}}{D _{q j'}} ,
      \end{equation}
where
\begin{equation}
      \label{app:5}
      D_{qj} = 1 - r^q _{j/0} r^q _{j/n} e^{2 i \beta _j d _j}
\end{equation}
\mbox{($d_0$ $\!=$ $\!d_n$ $\!=$ $\!0$)}. Here,
\begin{equation}
      \label{app:6}
      \beta _j = \sqrt { k_j ^2 - k^2}
      = \beta _j ' + i \beta _j ''
      \qquad  ( \beta_j ', \beta_j '' \geq 0 )
\end{equation}
($k$ $\!=$ $|\mathbf{k}|$), where
\begin{equation}
      \label{app:7}
      k _j = \sqrt {\varepsilon _j (\omega)} \,\frac {\omega} {c}
      = k _j ' + i k _j ''
      \qquad  ( k _j ', k _j '' \geq 0 ),
      \end{equation}
and 
$t_{j/j'}$
and $r_{j/j'}$ are, respectively, the transmission and reflection
coefficients between the layers $j'$ and $j$.
Finally, the unit vectors $\mathbf{e}^{(j)} _{q\pm} (\mathbf{k})$
in Eqs.~(\ref{app:2}) and (\ref{app:3})
are the polarization unit vectors for transverse electric  
\mbox{($q$ $\!=$ $\!s$)}
and transverse magnetic \mbox{($q$ $\!=$ $\!p$)} waves,
\begin{eqnarray}
      \label{app:8}
      &
      \displaystyle
      \mathbf{e}^{(j)}_{s\pm}(\mathbf {k})
      =
      \frac{\mathbf{k}}{k} \times  \mathbf{e} _z ,
&
      \\[.5ex]
& \displaystyle
      \label{app:9}
      \mathbf{e}^{(j)}_{p\pm}(\mathbf{k})
      =
      \frac {1} {k_j} \left(\mp \beta _j\frac{\mathbf{k}}{k}
      + k \mathbf{e} _z \right).
&
      \end{eqnarray}
In the limiting case where the space outside
the structure may be regarded as being vacuum 
the coefficients $c^+_\sigma ({\bf k},\omega)$, Eq.~(\ref{3.17}) read 
\begin{equation}
      \label{app:10}
      c^+_\sigma ({\bf k},\omega)
      = \frac{\pi \hbar}{\varepsilon _0}
     \frac{\omega^2}{c^2}
      \frac{1}{\beta_{0}}.
\end{equation}

\bibliography{bibl}

\end{document}